\documentclass[preprint]{aastex}

\shorttitle{Time delay in QSO 0957+561}
\shortauthors{Oscoz et al.}

\begin{document}

\title {Time delay in QSO 0957+561 from 1984-99 optical data\\
	}

\author{A. Oscoz\altaffilmark{1}, D. Alcalde\altaffilmark{1}, 
M. Serra--Ricart\altaffilmark{1},  
E. Mediavilla\altaffilmark{1}, C. Abajas\altaffilmark{1}, 
R. Barrena\altaffilmark{1}, J. Licandro\altaffilmark{1},
V. Motta\altaffilmark{1}, and J. A. Mu\~noz\altaffilmark{1}}
\altaffiltext{1}{
Instituto de Astrof\'\i sica de Canarias, E-38205 La Laguna, 
Tenerife, Spain}

\email{aoscoz@ll.iac.es, dalcalde@ll.iac.es, mserra@ot.iac.es, 
goicol@besaya.unican.es, emg@ll.iac.es, abajas@ll.iac.es, 
rbarrena@ll.iac.es, jlicandr@ll.iac.es, vmotta@ll.iac.es, 
jmunoz@ll.iac.es }

\begin{abstract}

Photometric optical data of QSO 0957+561 covering the period 
1984--99 are analyzed to discern between the two values of 
the time delay (417 and 424 days) mostly accepted in the 
recent literature. The observations, performed by groups 
from three different institutions---Princeton University, 
Harvard-Smithsonian Center for Astrophysics, and Ins\-tituto 
de Astrof\'\i sica de Canarias---and including new unpublished 
1998--9 data from the IAC80 Telescope, were obtained in five 
filters ($V$, $R$, $I$, $g$, and $r$). The different light 
curves have been divided into observational seasons and two 
restriction have been applied to better calculate the time 
delay: (i) points with a strange photometric behavior have 
been removed; and (ii) data sets without large gaps have been 
selected. Simulated data were generated to test several 
numerical methods intended to compute the time delay 
($\Delta\tau_{AB}$). The methods giving the best results---the 
discrete correlation function, $\delta$-square, z-transformed 
discrete correlation function, and linear interpolation---were 
then applied to real data. A first analysis of the 23 different 
time delays derived from each technique shows that 
$\Delta\tau_{AB}$ must be into the interval 420--424 days. 
From our statistical study, a most probable value of 
$\Delta\tau_{AB} = 422.6\pm0.6$ days is inferred.

\end{abstract}


\section{Introduction} \label{intro}

The first discovered gravitational lens system, QSO 0957+561 
(Walsh, Carswell, \& Weymann 1979), has been the subject of a 
continuous and exhaustive monitoring in several bands since 
1979. The special characteristics of this system made it very 
attractive for time delay determinations, and different values 
for this quantity were presented during the 1980s: $\Delta\tau = 
566\pm37$ days (Florentin-Nielsen 1984); $\Delta\tau = 376\pm37$ 
days (Schild \& Cholfin 1986); $\Delta\tau = 657\pm73$ days 
(Gondhalekar et al. 1986); $\Delta\tau = 478\pm73$ days (Leh\'ar, 
Hewitt, \& Roberts 1989). As can be seen, there was wide 
dispersion in the results obtained by different groups. However, 
the monitoring campaigns carried out during the early 1990s led 
to quite an odd situation, as all the results concentrate around 
two different values for the time delay: $\Delta\tau \sim 420$ 
days and $\Delta\tau \sim 510$ days. Calculations leading to the 
first value were presented by Vanderriest et al. (1989, 
$\Delta\tau = 415\pm20$ days); Schild (1990, $\Delta\tau = 
404\pm10$ days); and Pelt et al. (1994, $\Delta\tau = 415\pm32$ 
days in the $B$ band and $\Delta\tau = 409\pm23$ days in radio). 
On the other hand, a value close to 510 days was obtained by 
Beskin \& Oknyanskij (1992, $\Delta\tau = 522\pm15$ days in the 
$B$ band and $\Delta\tau = 515\pm15$ days in the $r$ band); 
Roberts et al. (1991, $\Delta\tau = 515\pm37$ days); and Press, 
Rybicki, \& Hewitt (1992a, $\Delta\tau = 536^{+14}_{-12}$ days 
in the $B$ band; 1992b, $\Delta\tau = 540\pm12$ days in $B$+radio). 

This situation abruptly changed when Kundi\'c et al. (1995) 
presented their observations in the $g$ and $r$ bands. A sharp 
drop appeared in 1994 December and could be used to discern 
between the ``long" (510 days) and the ``short" (420 days) time 
delay, provided continuous monitoring of QSO 0957+561 was 
carried out in the first half of 1996. This monitoring was 
performed, and the long time delay was rejected (Oscoz et al. 
1996; Kundi\'c et al. 1997). The controversy regarding the time 
delay seemed to be finally solved. However, the results 
appearing in the literature since 1995 concentrate again around 
two values, 417 and 424 days. These results are summarized as 
follows: $\Delta\tau = 423\pm6$ days (Pelt et al. 1996); 
$\Delta\tau = 417\pm3$ days (Kundi\'c et al. 1997); $\Delta\tau 
= 424\pm3$ days (Oscoz et al. 1997); $\Delta\tau = 425\pm17$ 
(Pijpers 1997); $\Delta\tau = 416.3\pm1.7$ days (Pelt et al. 
1998); $\Delta\tau = 425\pm4$ days (Serra-Ricart et al. 1999); 
$\Delta\tau = 417.4$ days (Colley \& Schild 2000). 

This difference is irrelevant in the Hubble constant 
calculations, as the uncertainty introduced by the time delay 
is much lower than the uncertainty given by other factors (e.g., 
the main lens galaxy's velocity dispersion or the lens modeling). 
However, the most accurate time delay should be used in the 
search for possible very rapid microlensing events in QSO 
0957+561, and a week's difference in $\Delta\tau$ could lead to 
the detection of false events or failure to detect real ones.

In this paper we have compiled photometric data of QSO 0957+561 
from three different observing groups covering the period 
1984--99 to obtain an estimate of the time delay by means of 
several statistical methods. This includes new unpublished data 
corresponding to the last observational campaign (1998--9) at 
the IAC80 Telescope. The data sets are presented in \S 2, 
while the methods for obtaining the time delay appear in \S 3. 
A first check of the goodness of these methods applied to 
simulated data is calculated in \S 4, and the best techniques 
are applied in \S 5 to real data. Finally, a discussion of our 
results appears in Section 6.

\section{Selected data sets}

Several monitoring campaigns of QSO 0957+561 in different bands 
have been performed since 1979. However, to obtain the time 
delay, only the observations obtained by groups from three 
different institutions will be considered here: Princeton 
University (hereafter PU), Harvard-Smithsonian Center for 
Astrophysics (hereafter CfA), and the Instituto de 
Astrof\'\i sica de Canarias (hereafter IAC).

\subsection{PU data}

Images were obtained at the Apache Point Observatory 3.5 m 
telescope in the $g$ and $r$ bands. QSO 0957+561 was monitored 
during several observational campaigns, although only data 
corresponding to the first two seasons have been published 
(Kundi\'c et al. 1995, 1997): (i) from 1994 December to 1995 
May; and (ii) from late 1995 to mid 1996. The resulting data 
comprise 51+46 $g$-band points and 54+46 $r$-band points. The 
light curves were calculated via aperture photometry, and have 
neither large error bars nor significant gaps. Their main 
characteristic is the presence of a sharp drop of about 0.1 mag 
in the $A$-component in late 1994 December, very useful for 
time-delay calculations.

\subsection{CfA data}

This data set is the largest ever obtained for a gravitational 
lens system. It consists of 1069 brightness measurements in 
the $R$ band, from late 1979 to mid 1996. The observations 
corresponding to the period 1979--89 were obtained at the 
Whipple Observatory 0.61 m telescope, while the remaining data 
were obtained with a 1.2 m telescope (Schild \& Thomson 1995, 
and references therein). The reduction procedure followed a 
basic aperture photometry scheme (although a new automated 
photometry reduction code is now being applied by the authors, 
the results do not substantially differ from the ``old" 
photometry). The error bars are not large, with the exception 
of the first five years. The main problem with this data set is 
the scarcity of observations during the first 1800 days (81 
brightness measurements). Moreover, those points coincide with 
the largest error bars. So, we will consider the data from mid 
1984 for time delay calculations.

\subsection{IAC data}

Lens monitoring was performed in four consecutive seasons 
(1996 February to June, 1996 October to 1997 July, 1997 
October to 1998 July, and 1998 October to 1999 July) using the 
CCD camera of the 82 cm IAC-80 Telescope (IAC80 hereafter), 
sited at the IAC's Teide Observatory (Tenerife, Canary Islands, 
Spain). A Thomson 1024$\times$1024 chip was used, offering a 
field of nearly 7$\farcm$5. Standard $VRI$ broad-band filters 
were used for the observations, corresponding fairly closely 
to the Landolt system (Landolt 1992). The final data set 
comprises 172 point in the $V$ band, 301 points in the $R$ 
band, and 112 points in the $I$ band. Accurate photometry was 
obtained by simultaneously fitting a stellar two-dimensional 
profile on each component by means of DAOPHOT software (see 
details in Serra-Ricart et al. 1999). A new, completely 
automatic IRAF task has been developed demonstrating, using a 
sample of simulated data, that the proposed method can achieve 
high-precision photometry. However, the errors bars obtained 
for IAC80 data are slightly larger than those of PU and CfA 
data. This could be explained by a decrease in chip sensibility 
due to the age of the CCD. In order to assess the reliability 
of our method using real data, simultaneous observations of 
QSO 0957+561 were undertaken on 1999 February 19 by using the 
IAC80 and the 2.5 m Nordic Optical Telescope (NOT hereafter) 
sited at the IAC's Roque de los Muchachos Observatory (La 
Palma, Canary Islands, Spain). The final reduced results are 
presented in Figure 1 and Table 1 (photometric errors for 
comparison star H and D are also included). The NOT light-curve 
errors (a few millimagnitudes) are much lower than the IAC80 
ones, and this difference could be explained in terms of the 
following: i) the NOT has a larger aperture than the IAC80, 
and ii) the NOT also has a better CCD chip. However, the good 
agreement between the two curves demonstrates that our 
reduction method works with high degree of accuracy. The 
photometric data are available at URL http://www.iac.es/lent.

\section{Different methods to obtain time delay estimates}

The large amount of data described in \S 2 adds new biases 
(different telescopes, filters, reduction processes, and 
behavior) to the inherent difficulty to analyse 
discrete, unevenly sampled temporary series. These facts leaded 
us to employ several statistical methods to calculate the time 
delay to increase the robustness of the results thus obtained. 
As a first step, several techniques will be checked by using 
simulated data: the discrete correlation function, dispersion 
spectra, $\delta^2$, $\delta^2$ modified, linear interpolation, 
and the $z$-transformed discrete correlation function. 

\subsection{Discrete correlation function (DCF)}

The DCF (Edelson \& Krolik 1988) is a technique valid for any
physical quantity that is observed to vary in time. For two 
discrete data trains, $A_i$ and $B_j$, the formula representing 
their DCF is
\begin{equation}
DCF(\tau) = {1 \over M} \, {\left( A_i - \bar A \right) \, \left( 
B_j - \bar B \right) \over {\sqrt{ \left( \sigma_A^2 - \epsilon_A^2 
\right) \, \left( \sigma_B^2 - \epsilon_B^2 \right)}}} \, ,
\end{equation}
averaging over the $M$ pairs for which $\tau - \alpha \leq
\Delta t_{ij} < \tau + \alpha$, $\alpha$, $\epsilon_k$, and 
$\sigma_f$ being the bin semi-size, the measurement error 
associated with the data set $k$, and the standard deviation, 
respectively. The maximum of the DCF is identified with the 
time delay.

\subsection{Dispersion spectra}

The data model (Pelt et al. 1996) consists of two time series, 
$A_i = q(t_i) + \epsilon_A (t_i)$, $i = 1,\dots,N_A$, and 
$B_j = q(t_j - \Delta\tau_{BA}) + l(t_j) + \epsilon_B (t_j)$, $j 
= 1,\dots,N_B$, where $q(t)$ represents the intrinsic variability 
of the quasar, $l(t)$ accounts for the difference in magnitudes 
plus additional variability in time due to microlensing, and 
$\epsilon_A (t)$ and $\epsilon_B(t)$ are observational errors. 
This two series are combined into one, $C$, for every fixed 
combination [$\tau$, $l(t)$] by taking all values of $A$ as they 
are and correcting the $B$ data by $l(t)$ and shifting them by 
$\tau$. The dispersion spectra that will be used here are  
represented by
\begin{equation}
D_{4,k}^2 = \min_{l(t)} {\sum_{n=1}^{N-1} \sum_{m=n+1}^{N} 
S_{n,m}^{(k)} W_{n,m} G_{n,m} (C_n - C_m )^2 \over  
\sum_{n=1}^{N-1} \sum_{m=n+1}^{N} S_{n,m}^{(k)} W_{n,m} G_{n,m}} 
\, ,
\end{equation}
where the $W_{n,m}$ are statistical weights of the combined light 
curves, and $G_{n,m} =1$ when $C_n$ and $C_m$ come from different 
curves ($A$ or $B$) and 0 otherwise. From eq. (2), we can consider 
two different approximations depending on the definition of 
$S_{n,m}^{(k)}$: (i) $S_{n,m}^{(1)} = 1$ if $|t_n - t_m | \leq 
\delta$ and 0 otherwise; and (ii) $S_{n,m}^{(2)} = 1 - |t_n - t_m | 
/ \delta$ if $|t_n - t_m | \leq \delta$ and 0 otherwise, $\delta$ 
being the maximum distance between two observations which can be 
considered as nearby. The minimum value of eq. (2) is assumed 
as the time delay.

\subsection{\boldmath $\delta$-square}

The $\delta^2$ method (Serra-Ricart et al. 1999) makes use of the
similarity between the discrete autocorrelation function (DAC) of 
the light curve of one of the components and the $A$--$B$ discrete
correlation function (DCF). From the DAC and DCF functions, 
one can define a function
\begin{equation}
\delta^2_m (\theta) = \left( {1 \over N} \right) \sum_{i = 1}^N 
S_i \left[ {\rm DCF}(\tau_i) - {\rm DAC}(\tau_i - \theta)
\right]^2 
\end{equation}
for every fixed value $\theta$ (days), where $S_i = 1$ when both 
the DCF and DAC are defined at $\tau_i$ and $\tau_i - \theta$, 
respectively, and 0 otherwise. The most probable value for the 
time delay should correspond to the minimum of this function.

\subsection{\boldmath $\delta^2$ modified}

This modification of the $\delta^2$ method was suggested by 
Schild (1999, private communication). It consists in comparing 
the DAC and DCF curves by taking their ratio instead of by 
calculating their difference. So, the final equation to obtain 
the time delay is
\begin{equation}
\delta^2_m (\theta) = \left( {1 \over N} \right) \sum_{i = 1}^N 
S_i \left[ {{\rm DCF}(\tau_i) \over {\rm DAC}(\tau_i - \theta)} 
\right]^2 \, .
\end{equation}

\subsection{Linear interpolation (LI)}

The linear method is similar to that suggested by Kundi\'c 
et al. (1997). One of the two light curves (hereafter light 
curve 1) is selected, and the linear interpolation of data and 
their errors is considered as reference. The other light curve 
(hereafter light curve 2) is then shifted in magnitude by just 
the difference between the means of both light curves. After 
that, light curve 2 is shifted in time and chi-square per 
degree of freedom ($\bar\chi^2$) for each time delay is 
calculated. The number of degrees of freedom is equal to the 
number of points of the light curve 2 in the overlapping 
interval minus 2 (because we are fitting shifts in magnitude 
and time). The time that minimizes $\bar\chi^2$ is taken as a 
provisional time delay.
 
This procedure is followed by using as reference both the $A$- 
and $B$-component light curves, selecting then the time delay 
closest to 421 days (the intermediate point between 417 and 
425 days). The uneven sampling of the light curves usually 
leads to a better time delay taking as a reference one of the 
two light curves. 
 
\subsection{\boldmath $z$-transformed discrete correlation 
function (ZDCF)}

The ZDCF (Alexander 1997) is a new method for estimating the
cross-correlation function (CCF) of sparse, unevenly sampled 
light curves. Fisher's $z$-transform of the linear correlation
coefficient, $r$, is used to estimate the confidence level of a
measured correlation. This technique attempts to correct the 
biases that affect the original DCF by using equal-population 
binning. The ZDCF involves three steps:

(i) All possible pairs of observations, $\{ a_i, b_j\}$, are sorted 
according to their time-lag, $t_i-t_j$, and binned into equal 
population bins of at least 11 pairs. Multiple occurrences of 
the same point in a bin are discarded so that each point appears 
only once per bin.

(ii) Each bin is assigned its mean time-lag and the intervals above
and below the mean that contain 1$\sigma$ of the points each.

(iii) The correlation coefficients of the bins are calculated and
$z$-transformed. The error is calculated in $z$-space and
transformed back to $r$-space. 

The time-lag corresponding to the maximum value of the ZDF is 
assumed as the time delay between both components.
 
\section{Analysis of the different methods by using simulated 
light curves} 

The application of the statistical methods described in \S 3 to 
simulated data sets can serve to check the validity of their 
results under different conditions, always with discrete and 
irregularly sampled data sets. The six selected data sets are 
quite similar to those presented in Serra-Ricart et al. (1999), 
where several sets of simulated photometric data with similar 
irregularity in the observations (time distribution of the 
data), magnitudes, and error bars (i.e., as large as or even 
larger than those of PU, CfA, and IAC; the worst situation is 
selected) to that of the IAC observations were created. 

First, a set of dates, $x_i$, between 1800 and 2000 (TJD = 
JD--2449000) approximately, was generated with a pseudo-random 
separation, taken from a uniform distribution between 0 and 5 
days. These data were alternatively separated in two time 
series, corresponding to $A$- and $B$-component light curves, 
this last curve being shifted by 420 days to simulate the 
existence of a time delay. A first magnitude was calculated 
for each date with the relationship $y_i = F(x_i)$ [see below 
for the different shapes of $F(x_i)$]. The probability of 
measuring a value $y$ for each $x_i$ is proportional to ${\rm 
e}^{[-(y-y_i)^2/2\sigma_i^2]}$, and hence characterized by 
$\sigma_i$, or, equivalently, the variable $d = y - y_i$ is 
distributed as ${\rm e}^{[-d^2/2\sigma_i^2]}$. A $\sigma_i$ 
taking pseudo-random values between 0.01 and 0.03 was generated 
for each $x_i$. From here the quantities $d_i$, pseudo-random 
numbers obtained from a normal Gaussian distribution with zero 
mean and standard deviation, $\sigma_i$, were calculated, 
allowing them to adopt positive or negative values. Finally, 
the magnitude was generated from the equation $y_o = F(x_i) + 
d_i = y_i + d_i$, with an error bar of $\sigma_i$. The $A$ 
component was made brighter by adding 0.1 to the magnitudes of 
the $B$ component. 

The first selected function was:
\begin{equation}
{\rm F1:} \;\; y = 17.17 + 0.5 \, {\rm e}^{-0.5 f} \, \sin (f), 
\,\,\;\; {\rm where} \;\; f = {\left( x - 1800 \right) \over 20} \, .
\end{equation}
This function represents light curves in which a sharp event 
similar to that reported by Kundi\'c et al. (1995) appears. 

The second function is
\begin{equation}
{\rm F2:} \;\; y = 17.2 + 0.1 \, \sin (f) \, \sin(4f),
\,\,\;\; {\rm where} \;\; f = {x \over 40} \, .
\end{equation}
In this case, the light curves present several maxima and minima,
although none of them is clearly remarkable.

An additional function, consistent with the actual variability 
of Q0957+561, was created (the IAC observational data from 
97--98 seasons, were selected as reference). The light curves 
were then fitted by the function
\begin{equation}
{\rm F3:} \;\; y = 17.07 -0.16 \, {\rm e}^f,
\,\,\;\; {\rm where} \;\; f={-(x - 15.8 - m)^2 \over 2 \, (10 +s)^2},
\end{equation}
$m$ being the mean of the TJD in the selected range and $s$ its 
standard deviation. The resulting simulated data show a lower 
variability to that obtained from $F1$ and $F2$.
 
Besides the comparison between the $A$- and $B$-component for 
the three data sets, an additional test was performed by 
removing some data of the $A$-component. By doing this, we want 
to simulate the data sets corresponding to certain periods in 
which no valid data could be obtained during several days (bad 
weather, problems with the telescope, etc.). 

The time delay corresponding to the different statistical 
techniques were first calculated by allowing their free 
parameters to vary. The best results were obtained with a 
$\alpha=5$ days for the DCF, $\alpha=2$ days for $D^2_{4,1}$ 
and $D^2_{4,2}$, and $\delta = 10$ days for $\delta^2$ and 
$\delta^2_m$. In the DCF, $\delta^2$ and $\delta^2_m$ cases, 
the results were quite similar for the parameters varying 
between 4 and 12 days. However, slightly better values were 
obtained for $\alpha=5$ days in the DCF case, and for $\delta 
= 10$ days in the $\delta^2$ and $\delta^2_m$ cases, so these 
quantities were selected. The objective in this paper is to 
analyze the different methods with several data sets by using 
the same conditions. An analysis of the best method and/or 
parameters to be used with a particular data set will be 
presented in a future paper. The uncertainties in the 
time-delay estimates were computed by generating 1000 
bootstrap samples and applying the statistical methods in each 
case.

Some interesting consequences can be derived from the results 
in Table 2. As expected, the best estimate of the time delay 
is always obtained for the function $F1$, i.e., when a sharp drop 
similar to the one appearing in \S 2.1 is present in the light 
curves. All the 
methods give a good value for the delay. On the other hand, in 
the $F2$ and $F3$ cases the error bar are too large for 
$\delta^2_m$, $D^2_{4,1}$, and $D^2_{4,2}$ when compared with 
the results obtained with the other methods. So only the DCF, 
$\delta^2$, ZDCF, and LI techniques will be applied to real data 
calculations, similar in most of the cases to the functions $F2$ 
and $F3$. Notice that the true time delay is always within the 
error bar, even when a large gap is present in the light curves. 
 
\section{Time delay from real data}
 
Prior to applying the different methods to calculate the time 
delay from real data, some considerations have to be taken 
into account. First, data should be checked to eliminate 
inconsistent measurements. This modification of the 
raw data, based on a suggestion by Falco (1997, private 
communication), takes account of the possible existence of 
strong and simultaneous (not time-shifted) variations of some 
data point in both components. The inclusion of such ``strange"
brightness records in the final data sets probably originated
from failures in the CCD or from bad weather conditions, 
creates artificial peaks or valleys in the light curve of one 
of the components. These maxima/minima have no importance when 
a sharp change in the behavior of the quasar is being analyzed, 
but can lead to a completely wrong time delay estimate when 
dealing with a smoother season. To avoid these abnormal 
observations we have removed the points with a simultaneous 
difference in magnitude in both components as compared with 
the previous and following records larger than 2.5 times their 
error bar. This was done by considering only those points with 
a difference in the observation dates of less than 10 days. 
The resulting data sets will be 
named bad point free (BPF). The 301 points of the IAC data set 
in the $R$ band are reduced to 289 with this restriction, while 
less than 60 of the 1069 CfA observations have to be removed. 
Finally, no brightness measurement of the PU data seems to be 
wrong. However, the BPF restriction could be applied neither 
to the IAC $I$- and $V$-band data nor to the CfA 89--90 and 
91--92 $R$-band data due to their low number of points, with a 
minimum distance between neighboring points more than 10 
days in most of the cases.

Another drawback when dealing with real data is the 
impossibility of observing the system during certain months of 
the year and the consequent lack of suitable edges. Once it is 
stated, see \S 1, that the rough value of the time delay is 
around 420 days, the comparison between the $A$ and $B$ 
components should be made by previously selecting a ``clean" 
data set (CD, see Serra-Ricart et al. 1999), i.e., homogeneous 
monitoring of both images during two active and clear (free 
from large gaps) epochs separated by $\sim 420$ days. 

Finally, both types of corrections, BPF and CD, were combined 
to obtain the definitive data sets used in this paper. To see 
the importance of applying the CD--BPF corrections, we mention 
two extreme examples: 1) $\Delta\tau = 398\pm11$ days with the 
raw IAC data corresponding to the 96--7 season, while 
$\Delta\tau = 428\pm9$ days with the CD--BPF approximation; and 
(2) $\Delta\tau = 384\pm5$ days with the original 93--4 CfA data 
and $\Delta\tau = 423\pm2$ days with the CD--BPF approximation.

To summarize, Monte Carlo calculations were applied to the four 
techniques (DCF, $\delta^2$, ZDCF, and LI) with the CD and the 
CD-BPF restrictions. The PU, IAC, and CfA data sets were divided 
into observational seasons, leading to the 23 different 
time-delay estimates per method appearing in Tables 3 to 6. The 
results obtained from the PU data in both the $g$ and $r$ filter
are presented in Table 3. Notice that two of the methods (DCF and 
LI) employed here are also used by Kundi\'c et al. (1997). Our 
results are quite similar to those obtained by these authors, and, 
moreover, they are always into their error. The small differences 
come from the selection of clean data sets. Tables 4 and 5 offer 
the time delay 
for the IAC data in the $R$, and in the $I$ and $V$ bands, 
respectively. Finally, Table 6 was obtained from the CfA $R$ band 
data. The analysis of these delays are done by considering 
the CD--BPF quantities, except in those cases in which only the 
CD results could be obtained. The quantities obtained for the 
CfA 92--3 season with the DCF and $\delta^2$ methods can be 
discarded, as they appear to be clearly inconsistent (394 $\pm$ 1 
and 403 $\pm$ 1 days, respectively). 

\section{Discussion and conclusions}

A first step in discussing the value of $\Delta\tau_{AB}$ 
consists in computing, for each of the techniques (DCF, 
$\delta^2$, ZDCF, and LI), the number of occurrences of each 
value of the time delay. These quantities, obtained from Tables 
3 to 6, are depicted by the black lines in Figure 2: DCF (top 
panel, left); $\delta^2$ (top panel, right); ZDCF (bottom panel, 
left); and LI (bottom panel, right). (The different values of 
$\Delta\tau_{AB}$ have been grouped into two-day bins). As can 
be seen, two remarkable characteristics can be deduced from 
Figure 2: (i) there is a small dispersion in the delays, as 
most of them are in the interval 415--428 days; and (ii) the 
centroids of the histograms, given by the average of the time 
delays derived in Tables 3 to 6, are always in the interval 
420--424 days. These last quantities are represented in Figure 
2 (open circles) together with their uncertainty (r.m.s., see 
Table 7 and discussion below). Note that the largest peak of 
each histogram coincides with this average value, except for the 
ZDCF technique. In the histogram corresponding to $\delta^2$, 
the maximum corresponds to the values 421--424, while the 
average is 421.8 $\pm$ 1.3 days. The DCF panel does not lead 
to a clear time delay, with two maxima in the intervals 
417--420 and 423--424 days. The average here is given by 
423.3 $\pm$ 1.4 days. In the case of LI, the peak is placed 
at $\Delta\tau_{AB}$ = 423--424 days, and the average is 
424.3 $\pm$ 1.2 days. Finally, the ZDCF method has a maximum 
around 423--424 days. However, the average is 420.6 $\pm$ 1.1 
days, which is slightly different. The red histograms in Figure 
2 represent the total number of occurrences obtained by adding 
the results from the four techniques. Its center is again over 
420 days, giving a maximum of 423--424 days.

To complement these calculations, which have been done by 
fixing the method and computing the probability of appearance 
of each delay, we can now represent the number of times each 
value appears for each data set, independently of the method 
employed. Four different data sets have been selected: (i) PU 
$r$ and $g$ filters, Figure 3a; (ii) IAC $R$ data, Figure 3b; 
(iii) IAC $V$ and $I$ records, Figure 3c; and (iv) CfA $R$ data, 
Figure 3d. In this case, the centroid of the distributions, 
represented by open circles in Figure 3, is again over 420 days: 
421.4 $\pm$ 1.1 days, 423.7 $\pm$ 1.3 days, 423.7 $\pm$ 
1.2 days, and 421.8 $\pm$ 1.0, for PU $r$ and $g$, IAC $R$, 
IAC $I$ and $V$, and CfA $R$, respectively. A first positive 
consequence of the PU results is their extremely short 
dispersion indicating the goodness of the data and the presence 
of the sharp event. The maximum here is placed between 419 and 
424 days. However, the clearest peak appears from the IAC $R$ 
values around 423--424 days. The IAC $I$ and $V$ panel shows 
more dispersion, probably due to the higher error bars of the 
light curves. There is not a unique maximum here, as two peaks
appear around 423--424 and 427--428 days. The highest 
dispersion in the results can be seen in Figure 3d, 
corresponding to the CfA data. The maximum would in this case 
be in the interval 417--422 days. A remarkable result here is 
that the average coincides with the maxima in all the panels. 
Once again, the red histogram gives the total number of 
occurrences.

The combination of the results derived from Figures 2 and 3 
(centroids and maxima of the distributions) support a 
$\Delta\tau_{AB}$ in the range of 420--424 days, although a 
time delay of around 417 days cannot be totally discarded. It 
is important to remark that the results are the same 
independently of the technique employed or of the data set 
selected. 

The time delay between A and B components of Q0957+561 does 
not depend on the filter (as it is an acromatic effect) and/or 
the time (different campaigns), so it should be possible to 
merge the different sample results. A very important point is 
to assess the statistical reliability of the different delay 
calculation methods in order to estimate final delay errors. 
Several statistics were calculated, Mean Delay ($MD = 
\sum_{i=1}^{N}\Delta\tau_{ABi} / N$), Mean Error ($ME = 
\sum_{i=1}^{N} \epsilon_i / N$, with $\epsilon_i$ individual 
errors), and Dispersion ($DI = 
[\sum_{i=1}^{N}(\Delta\tau_{ABi} - MD)^2 / (N-1)]^{1/2}$). If 
the error estimate is correct, then $ME \approx DI$, and the 
final error for the time delay will be given by the r.m.s. 
($[\sum_{i=1}^{N}(\Delta\tau_{ABi} - MD)^2 / (N*(N-1))]^{1/2}$) 
(see Eadie et al. 1971). Tables 7 to 10 (see below) show the 
final results for the four methods. In all cases, within the 
statistical errors, good agreement is found between the mean 
error and the dispersion.

When this procedure is applied to the different time delays 
given by each technique, one obtains the results in the first 
four rows of Table 7, where the mean values of the time delay 
and the uncertainties (r.m.s.) are shown. According to these 
calculations, the definitive time delay would be in the 
interval 420(ZDCF)--424(LI) days, with an uncertainty below 
1.4 days. This interval coincides with that derived from the 
analysis of Figures 2 and 3. When this calculation is done with 
only the $R$ band results (Table 8), the results are almost 
identical.

The different analyses performed until now have been done 
considering all the results obtained in Tables 3 to 6. However, 
the error bars of some of these values exclude the interval 
420--424 days, where, as shown before, there is the highest 
probability of finding the right $\Delta\tau_{AB}$. The mean 
delays and uncertainties obtained when these values are removed 
(first four rows of Table 9) are very similar to those of Table 7. 
$\Delta\tau_{AB}$ is now restricted to the interval 420.8--423.2 
days, with an uncertainty below 1.3 days. Once again, the 
results derived from the $R$-filter data are almost the same 
(Table 10).

The validity of these statistical results leaded us to repeat 
the calculations of Tables 7 to 10, but this time having into 
account all the delays, i.e., without considering the method
(DCF, $\delta^2$, ZDCF, or LI). This allows us to have four 
different time delays per year in most of the occasions, and 
so, a larger amount of data for the statistical analysis. The 
results appear in the last row of Tables 7 to 10, although 
only the values of Tables 7 and 9, $\Delta\tau_{AB} = 
422.6\pm0.6$ days and $\Delta\tau_{AB} = 422.0\pm0.6$ days, 
respectively, will be considered, as they were obtained from 
a larger amount of data. Adopting a conservative point of view, 
we will select the quantity with the higher uncertainty, 
$\Delta\tau_{AB} = 422.6\pm0.6$ days, as the final time delay.

Different treatments of the time delays 
obtained in \S 5 have been performed. The analysis of these 
results always points to a time delay in the interval 
420--424 days. None of these methods clearly favors the values 
416--418 days as the right time delay. Moreover, the averages 
of the quantities of Tables 7 and 9 give values around 422 days, 
coinciding with the maxima obtained in Figures 2 and 3. Assuming 
this value as the time delay between the $A$ and $B$ components of 
Q0957+561, let us check which of the results given in Tables 3 
to 6 include 422 days in their the error bars. Figure 4 
offers the number of times, written as a percentage of the total 
number of time delays obtained for each method, that each value 
of $\Delta\tau_{AB}$ is included in these error bars. This 
probability is the same, 86\%, for DCF, $\delta^2$ and ZDCF, 
and 74\% for LI. In this case, the DCF and $\delta^2$ methods 
show the highest probability for a $\Delta\tau_{AB}$ of 421--422 
days, while the peak in the LI curve is placed at 423 days. A 
wider maximum is obtained in the case of ZDCF, with the same 
probability between 418 and 421 days. 

As can be seen, not all the seasons in the different data sets 
are fully appropriate for calculating the time delay. However, our 
intention was to analyze all the data in the three different 
data sets with four different methods, and finally to restrict 
the uncertainty in the calculation of the time delay. The 
statistical treatment of all the results confirms a time delay 
of $\approx 422$ days.

\acknowledgements

We are especially grateful to E. E. Falco for advising us on 
the possible presence of strange points in our data sets, to R. 
Schild for helpful comments on the statistical methods, and to T. 
Alexander for providing us with his programs to calculate the 
ZDCF method and his help in understanding it. This work was 
supported by the P6--88 project of the Instituto de 
Astrof\'\i sica de Canarias (IAC), Universidad de Cantabria 
funds, and DGESIC (Spain) grant PB97-0220-C02.

%
%

\clearpage

%
%

\figcaption[fig1a.eps]{Light curves for $A$ and $B$ images of 
QSO 0957+561 in the $R$ band obtained on the night of 1999 
February 19. IAC80 data are represented by boxes and NOT data 
are indicated by circles. One-sigma error bars are indicated. 
See text for more details.}

\figcaption[dels2e.eps]{Number of times (black lines) that each 
value of the time delay appears, obtained from the results of 
Tables 3 to 6. The four different techniques have been 
considered: a) DCF, b) $\delta^2$, c) ZDCF, d) LI. The open 
circles represent the average value of the delays for each 
method, and the red 
histograms represent the sum of the values given by each method.}

\figcaption[dels5b.eps]{Number of times (black lines) that each 
value of the time delay appears for each data set from the
values given in Tables 3 to 6. Four data sets are represented: 
a) PU $r$ and $g$, b) IAC $R$, c) IAC $I$ and $V$, and d) CfA $R$. 
The open circles represent the average value of the delays for 
each data set, and the red histograms represent the total sum 
of the values.}

\figcaption[dels3c.eps]{Percentage of times that a value of the 
time delay is included in the error bars of the results given 
in Tables 3 to 6: black = DCF, red = $\delta^2$, green = ZDCF, 
and blue = LI.}


%
%

\clearpage
\begin{figure*}[h]
\plotone{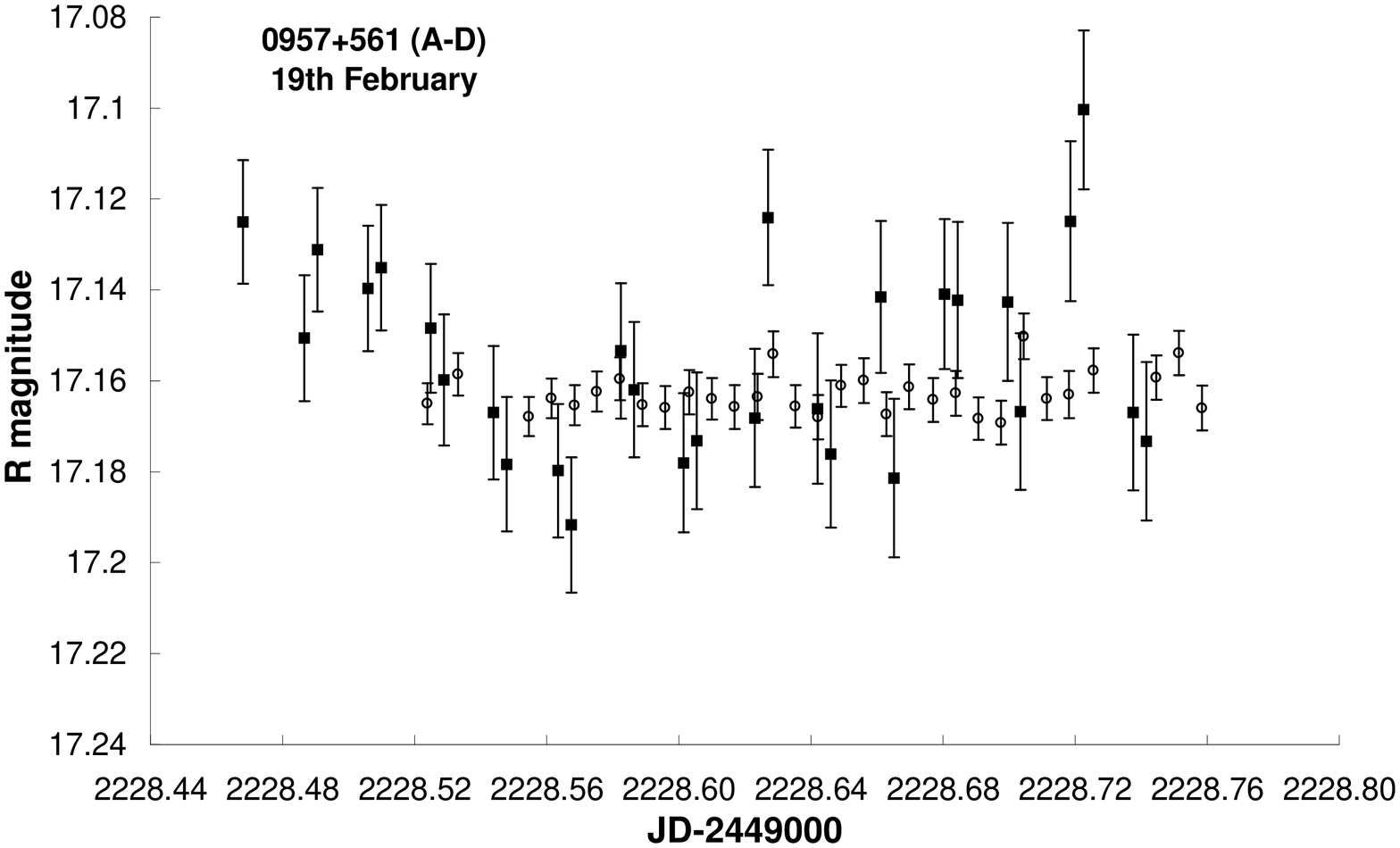}
\plotone{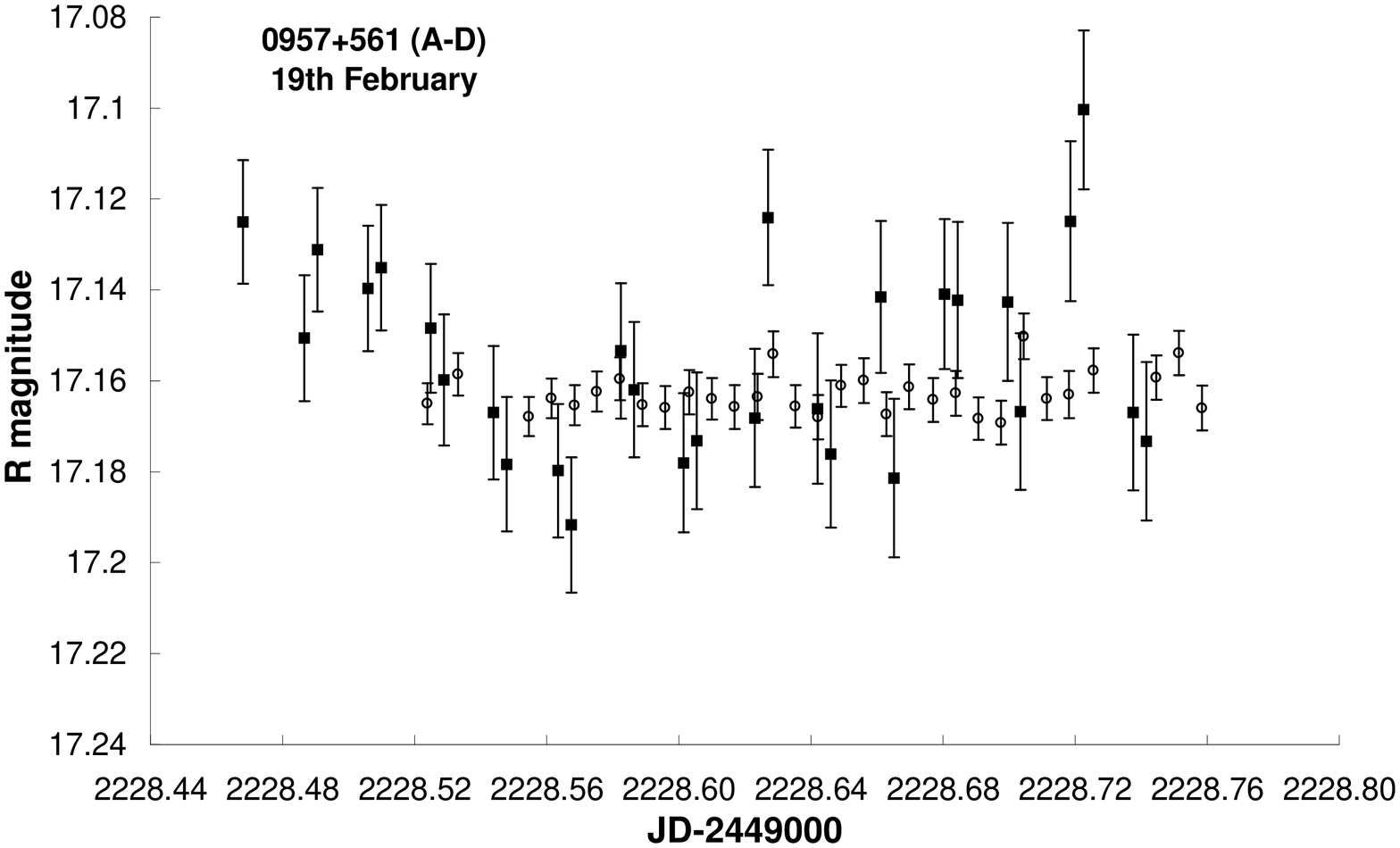}
Fig. 1
\end{figure*}

\clearpage
\begin{figure*}[h]
\plotone{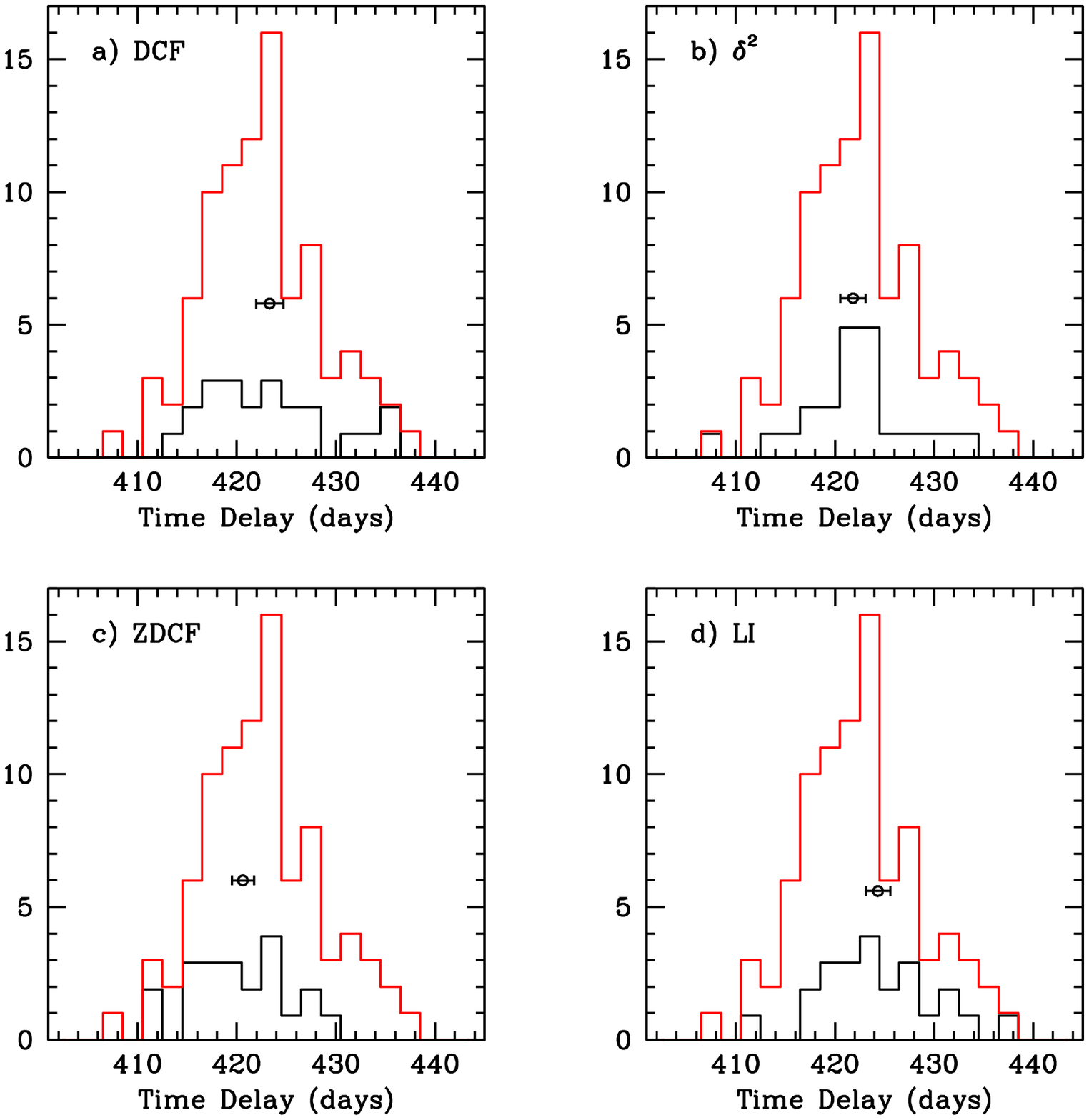}
Fig. 2
\end{figure*}

\clearpage
\begin{figure*}[h]
\plotone{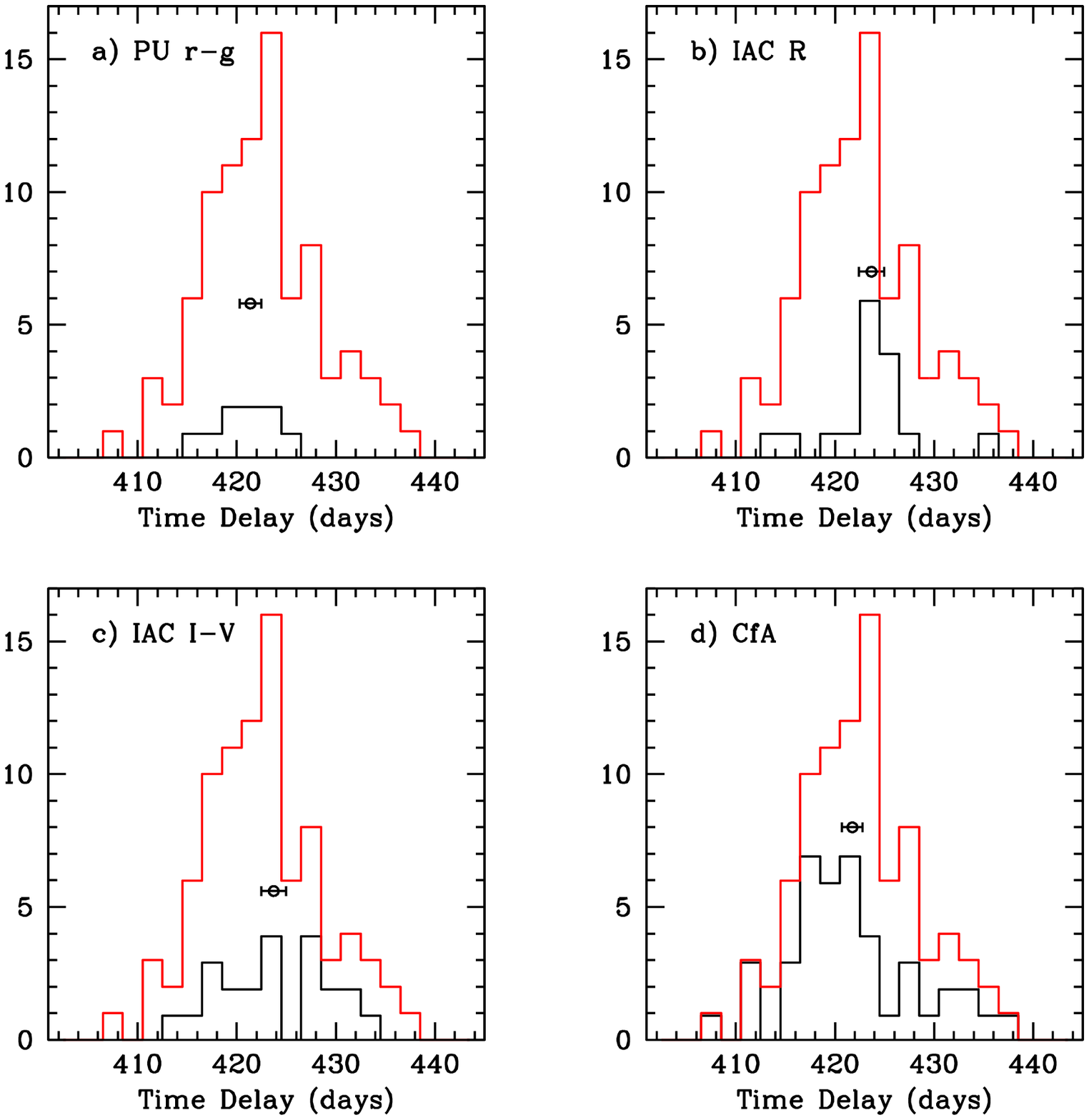}
Fig. 3
\end{figure*}

\clearpage
\begin{figure*}[h]
\plotone{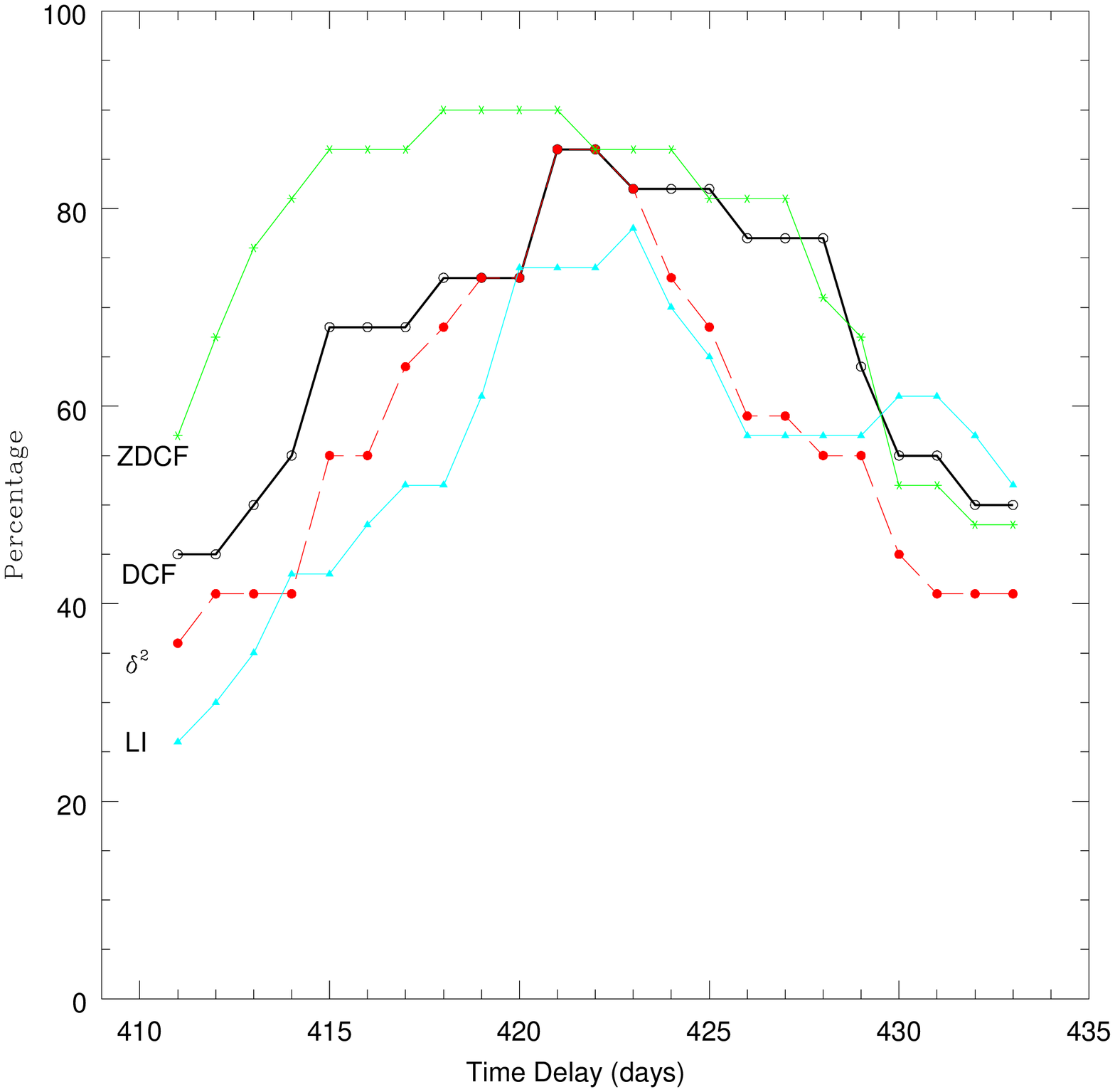}
Fig. 4
\end{figure*}

%
%

\clearpage
\begin{deluxetable}{ccc}
\tablecaption{$R$ photometric errors}
\tablewidth{0pt}
\tablehead{
\colhead{Light curve\tablenotemark{a}}& \colhead{NOT\tablenotemark{b}} &\colhead{IAC80}}
\startdata
A   & 0.005 & 0.016  \\
B   & 0.005 & 0.014  \\
H-D & 0.002 & 0.005  \\
\enddata
\tablenotetext{a}{See section 1 for details.}
\tablenotetext{b}{Error values are expressed in magnitudes.}
\end{deluxetable}

\clearpage
\begin{table}
\begin{center}
\caption{``Time delays" obtained from the application of a Monte 
Carlo algorithm with the different techniques to the six simulated 
data sets (see main text for details). The ``true" time delay 
was 420 days.}
\begin{tabular}{lcccccc}
\tableline
\tableline
Method&$F1$&$F1$ (gap in $A$)&$F2$&$F2$ (gap in $A$)&$F3$&$F3$ (gap 
in $A$)\\
\tableline
DCF ($\alpha =
5$)&421$\pm$2&421$\pm$4&419$\pm$1&420$\pm$2&415$\pm$9&412$\pm$9\\
$\delta^2$ ($\delta =
10$)&422$\pm$2&421$\pm$2&419$\pm$1&420$\pm$1&420$\pm$5&417$\pm$5\\
ZDCF&
420$\pm$3&420$\pm$3&421$\pm$1&420$\pm$7&418$\pm$14&416$\pm$13\\
LI&421$\pm$2&421$\pm$2&420$\pm$1&421$\pm$1&419$\pm$12&418$\pm$13\\
$\delta^2_m$ ($\delta = 10$)&
422$\pm$9&422$\pm$13&425$\pm$30&416$\pm$34&403$\pm$34&413$\pm$29\\
$D^2_{4,1}$ ($\alpha =
2$)&421$\pm$3&421$\pm$4&421$\pm$12&416$\pm$20&422$\pm$27&414$\pm$25\\
$D^2_{4,2}$ ($\alpha =
2$)&421$\pm$2&421$\pm$3&423$\pm$15&415$\pm$26&420$\pm$26&412$\pm$21\\
\tableline
\end{tabular}
\end{center}
\end{table}

\clearpage
\begin{deluxetable}{ccccc}
\tablecaption{Time delays obtained from the application of a 
Monte Carlo algorithm with four different methods to the PU 
$r$ and $g$ data}
\tablewidth{0pt}
\tablehead{
\colhead{Filter} & \colhead{Years} & \colhead{Method} & 
\colhead{Points} & \colhead{Delay}\\
\colhead{}&\colhead{}&\colhead{}&\colhead{(CD)}&\colhead{(CD)}}
\startdata
&&DCF&&422$\pm$1\\
{\it g}& $A$ (94--5); $B$ (95--6)& $\delta^2$& 42; 39& 423$\pm$1\\
&&ZDCF&&416$\pm$2\\
&&LI&&419$\pm$5\\
\tableline
&&DCF&&426$\pm$5\\
{\it r}& $A$ (94--5); $B$ (95--6)& $\delta^2$& 41; 41& 423$\pm$2\\
&&ZDCF&&420$\pm$8\\
&&LI&&422$\pm$3\\
\enddata
\end{deluxetable}

\clearpage
\begin{deluxetable}{cccccc}
\tablecaption{Time delays obtained from the application of a 
Monte Carlo algorithm with four different methods to the IAC 
$R$ data}
\tablewidth{0pt}
\tablehead{
\colhead{Years} & \colhead{Method} & \colhead{Points} & 
\colhead{Delay} & \colhead{Points} & \colhead{Delay}\\
\colhead{} & \colhead{} & \colhead{(CD)}& \colhead{(CD)} & 
\colhead{(CD-BPF)}& \colhead{(CD-BPF)}}
\startdata
&DCF&&428$\pm$12&&425$\pm$11\\
$A$ (96--8); $B$ (97--9)&$\delta^2$&182; 220&421$\pm$5&173; 212&421$\pm$4\\
&ZDCF&&418$\pm$13&&426$\pm$13\\
&LI&&421$\pm$16&&424$\pm$17\\
\tableline
&DCF&&432$\pm$8&&436$\pm$8\\
$A$ (96); $B$ (96--7)&$\delta^2$&31; 36&425$\pm$8&28; 33&428$\pm$9\\
&ZDCF&&422$\pm$13&&424$\pm$14\\
&LI&&424$\pm$7&&424$\pm$7\\
\tableline
&DCF&&436$\pm$10&&424$\pm$13\\
$A$ (96--7); $B$ (97--8)&$\delta^2$&46; 86&431$\pm$4&45; 84&425$\pm$4\\
&ZDCF&&429$\pm$13&&424$\pm$12\\
&LI&&415$\pm$16&&426$\pm$16\\
\tableline
&DCF&&414$\pm$12&&416$\pm$12\\
$A$ (97--8); $B$ (98--9)&$\delta^2$&79; 62&414$\pm$10&78; 61&413$\pm$8\\
&ZDCF&&420$\pm$12&&420$\pm$13\\
&LI&&427$\pm$13&&423$\pm$13\\
\enddata
\end{deluxetable}

\clearpage
\begin{deluxetable}{ccccc}
\tabletypesize{\footnotesize}
\tablecaption{Time delays obtained from the application of a 
Monte Carlo algorithm with four different methods to the IAC 
$I$ and $V$ data}
\tablewidth{0pt}
\tablehead{
\colhead{Filter} & \colhead{Years} & \colhead{Method} & 
\colhead{Points} & \colhead{Delay}\\
\colhead{}&\colhead{}&\colhead{}&\colhead{(CD)}&\colhead{(CD)}}
\startdata
&&DCF&&415$\pm$13\\
{\it I}&$A$ (96--8); $B$ (97--9)&$\delta^2$&65; 87&423$\pm$12\\
&&ZDCF&&417$\pm$14\\
&&LI&&419$\pm$9\\
\tableline
&&DCF&&424$\pm$16\\
{\it I}&$A$ (96); $B$ (96--7)&$\delta^2$&19; 18&424$\pm$16\\
&&ZDCF&& \\
&&LI&&430$\pm$3\\
\tableline
&&DCF&&419$\pm$9\\
{\it I}&$A$ (96--7); $B$ (97--8)&$\delta^2$&18; 31&422$\pm$7\\
&&ZDCF&&\\
&&LI&&427$\pm$7\\
\tableline
&&DCF&&414$\pm$15\\
{\it I}&$A$ (97--8); $B$ (98--9)&$\delta^2$&28; 39&417$\pm$18\\
&&ZDCF&&421$\pm$14\\
&&LI&&427$\pm$11\\
\tableline
&&DCF&&432$\pm$11\\
{\it V}&$A$ (96--8); $B$ (97--9)&$\delta^2$&68; 149&429$\pm$11\\
&&ZDCF&&418$\pm$9\\
&&LI&&428$\pm$14\\
\tableline
&&DCF&&433$\pm$7\\
{\it V}&$A$ (97--8); $B$ (98--9)&$\delta^2$&29; 84&432$\pm$6\\
&&ZDCF&&428$\pm$9\\
&&LI&&423$\pm$3\\
\enddata
\end{deluxetable}

\clearpage

\begin{deluxetable}{cccccc}
\tabletypesize{\scriptsize}
\tablecaption{Time delays obtained from the application of a 
Monte Carlo algorithm with four different methods to the CfA 
$R$ data}
\tablewidth{0pt}
\tablehead{
\colhead{Years} & \colhead{Method} & \colhead{Points} & 
\colhead{Delay} & \colhead{Points} & \colhead{Delay}\\
\colhead{} & \colhead{} & \colhead{(CD)}& \colhead{(CD)} & 
\colhead{(CD-BPF)}& \colhead{(CD-BPF)}}
\startdata
&DCF&&420$\pm$6&&419$\pm$4\\
$A$ (84--5); $B$ (85--6)&$\delta^2$&40; 49&420$\pm$5 &39; 47&421$\pm$6\\
&ZDCF&&429$\pm$7 &&428$\pm$7\\
&LI&&432$\pm$5 &&431$\pm$6\\
\tableline
&DCF&&419$\pm$10&&417$\pm$12\\
$A$ (85--6); $B$ (86--7)&$\delta^2$&49; 74&410$\pm$13 &46; 72&407$\pm$15\\
&ZDCF&&420$\pm$9 &&417$\pm$12\\
&LI&&409$\pm$14 &&431$\pm$8\\
\tableline
&DCF&&431$\pm$12&&428$\pm$13\\
$A$ (86--7); $B$ (87--8)&$\delta^2$&53; 58&434$\pm$10 &52; 58&418$\pm$17\\
&ZDCF&&429$\pm$15 &&430$\pm$12\\
&LI&&424$\pm$9 &&420$\pm$5\\
\tableline
&DCF&&425$\pm$12 &&428$\pm$13\\
$A$ (87--8); $B$ (88--9)&$\delta^2$&60; 53&423$\pm$17&58; 53&422$\pm$18\\
&ZDCF&&419$\pm$17 &&411$\pm$13\\
&LI&&421$\pm$11 &&422$\pm$12\\
\tableline
&DCF&&417$\pm$21 &&\\
$A$ (88--9); $B$ (89--90)&$\delta^2$&23; 19&420$\pm$10& &\\
&ZDCF&&412$\pm$8 &&\\
&LI&&421$\pm$2 &&\\
\tableline
&DCF&&440$\pm$5 &&424$\pm$19\\
$A$ (89--90); $B$ (90--1)&$\delta^2$&40; 38&407$\pm$7&40; 34&420$\pm$3\\
&ZDCF&&417$\pm$6 &&420$\pm$7\\
&LI&&410$\pm$6 &&411$\pm$3\\
\tableline
&DCF&&435$\pm$8 &&\\
$A$ (90--1); $B$ (91--2)&$\delta^2$&15; 29&433$\pm$18& &\\
&ZDCF&&423$\pm$19 &&\\
&LI&&426$\pm$6 &&\\
\tableline
&DCF&& 394$\pm$0&&394$\pm$1\\
$A$ (91--2); $B$ (92--3)&$\delta^2$&14; 72&402$\pm$1&14; 67&403$\pm$1\\
&ZDCF&&423$\pm$12 &&416$\pm$13\\
&LI&&436$\pm$9 &&437$\pm$8\\
\tableline
&DCF&&411$\pm$5 &&419$\pm$6\\
$A$ (92--3); $B$ (93--4)&$\delta^2$&70; 98&423$\pm$2&63; 93&423$\pm$2\\
&ZDCF&&411$\pm$11 &&424$\pm$15\\
&LI&&423$\pm$13 &&433$\pm$3\\
\tableline
&DCF&&422$\pm$4 &&422$\pm$4\\
$A$ (93--4); $B$ (94--5)&$\delta^2$&83; 111&421$\pm$2&78; 95&421$\pm$2\\
&ZDCF&&422$\pm$6 &&422$\pm$7\\
&LI&&421$\pm$4 &&418$\pm$6\\
\tableline
&DCF&&415$\pm$6 &&418$\pm$7\\
$A$ (94--5); $B$ (95--6)&$\delta^2$&101; 66&411$\pm$2&87; 57&415$\pm$3\\
&ZDCF&&416$\pm$2 &&416$\pm$5\\
&LI&&421$\pm$12 &&418$\pm$5\\
\enddata
\end{deluxetable}

\clearpage
\begin{deluxetable}{ccccc}
\tablecaption{Statistical results obtained from the 23 time delays 
calculated for each method}
\tablewidth{0pt}
\tablehead{
\colhead{Method} & \colhead{Mean Delay} & \colhead{Mean Error} & 
\colhead{Dispersion} & \colhead{r.m.s.}}
\startdata
DCF&423.3&10&7&1.4\\
$\delta^2$&421.8&9&6&1.3\\
ZDCF&420.6&11&5&1.1\\
LI&424.3&7&6&1.2\\
&&&&\\
Total&422.6&9&6&0.6\\
\enddata
\end{deluxetable}

\begin{deluxetable}{ccccc}
\tablecaption{As in Table 7 but only considering $R$ filter 
time delays}
\tablewidth{0pt}
\tablehead{
\colhead{Method} & \colhead{Mean Delay} & \colhead{Mean Error} & 
\colhead{Dispersion} & \colhead{r.m.s.}}
\startdata
DCF&423.4&11&6&1.7\\
$\delta^2$&420.5&8&6&1.7\\
ZDCF&420.9&11&6&1.4\\
LI&424.3&8&7&1.7\\
&&&&\\
Total&422.3&10&6&0.8\\
\enddata
\end{deluxetable}

\clearpage
\begin{deluxetable}{ccccc}
\tablecaption{Statistical results obtained only considering time 
delays and error bars than include part or the whole interval 
given by the results of Table 6 (420-424 days)}
\tablewidth{0pt}
\tablehead{
\colhead{Method} & \colhead{Mean Delay} & \colhead{Mean Error} & 
\colhead{Dispersion} & \colhead{r.m.s.}}
\startdata
DCF&422.2&11&6&1.3\\
$\delta^2$&421.6&9&6&1.2\\
ZDCF&420.8&11&5&1.2\\
LI&423.2&8&4&0.9\\
&&&&\\
Total&422.0&10&5&0.6\\
\enddata
\end{deluxetable}

\begin{deluxetable}{ccccc}
\tablecaption{As in Table 9 but only considering $R$ filter 
time delays}
\tablewidth{0pt}
\tablehead{
\colhead{Method} & \colhead{Mean Delay} & \colhead{Mean Error} & 
\colhead{Dispersion} & \colhead{r.m.s.}}
\startdata
DCF&422.5&11&6&1.6\\
$\delta^2$&420.9&9&6&1.8\\
ZDCF&420.9&11&6&1.4\\
LI&423.0&8&4&1.2\\
&&&&\\
Total&421.7&10&5&0.8\\
\enddata
\end{deluxetable}


\clearpage
\begin{thebibliography}{}
\bibitem[a]{} Alexander, T. 1997, Astronomical Time Series, Maoz, 
	D., Sternberg, A., and Leibowitz, E. M. (eds), Dordrecht: 
	Kluwer, 163
\bibitem[a]{} Beskin, G. M., and Oknyanskij, V. L. 1992, Lecture 
	Notes in Physics 406, Gravitational Lenses, Kayzer, R., 
	Schramm, T., and Refsdal S. (eds), Springer--Verlag: 
	Heidelberg, 67
\bibitem[a]{} Colley, W. N., and Schild, R. E. 2000, \apj, 
	540, 104
\bibitem[a]{} Eadie, W. T., Drijard, D., James, F. E., and 
	Roos, M. 1971, Statistical Methods in Experimental 
	Physics (Amsterdam: North-Holland)
\bibitem[a]{} Edelson, R. A., and Krolik, J. H. 1988, \apj, 333, 646
\bibitem[a]{} Florentin-Nielsen, R. 1984, A\&A, 138, 119
\bibitem[a]{} Gondhalekar, P. M., Wilson, R., Dupree, A. K., 
	and Burke, B.F. 1986, London Conference Proceedings, New 
	Insights in Astrophysics: Eight Years of Astronomy with 
	IUE, SP-263, 715
\bibitem[a]{} Kundi\'c, T., Colley, W. N., Gott III, J. R., 
	Malhotra, S., Pen, U., Rhoads, J. E., Stanek, K. Z., 
	Turner, E. L., and Wambsganss, J. 1995, \apjl, 455, L5
\bibitem[a]{} Kundi\'c, T., Turner, E. L., Colley, W. N., 
	Gott III, J. R., Rhoads, J. E., Wang, Y., Bergeron, L. E., 
	Gloria, K. A., Long, D. C, Malhotra, S., and Wambsganss, 
	J. 1997, \apj, 482, 75
\bibitem[a]{} Landolt, A. U. 1992, \aj, 104, 340
\bibitem[a]{} Leh\'ar, J., Hewitt, J. N., and Roberts, D. H. 
	1989, Gravitational Lenses, Moran, J. M., Hewitt, J. N., 
	and Lo, K. Y. (eds), Gravitational Lenses, Dordrecht: 
	Reidel, 84
\bibitem[a]{} Oscoz, A., Serra-Ricart, M., Goicoechea, L. J., 
	Buitrago J., and Mediavilla E. 1996, \apjl, 470, L19
\bibitem[a]{} Oscoz, A., Mediavilla, E., Goicoechea, L. J., 
	Serra-Ricart, M., and Buitrago, J. 1997, \apjl, 479, L89
\bibitem[a]{} Pelt, J., Hoff, W., Kayser, R., Refsdal, S., and 
	Schramm, T. 1994, A\&A, 256, 775
\bibitem[a]{} Pelt, J., Kayser, R., Refsdal, S., and Schramm, 
	T. 1996, A\&A, 305, 97
\bibitem[a]{} Pelt, J., Schild, R., Refsdal, S., and Stabell, 
	R. 1998, A\&A, 336, 829
\bibitem[a]{} Pijpers, F. P. 1997, \mnras, 289, 933
\bibitem[a]{} Press, W. H., Rybicki, G. B., and Hewitt, J. N. 
	1992a, \apj, 385, 404
\bibitem[a]{} Press, W. H., Rybicki, G. B., and Hewitt, J. N. 
	1992b, \apj, 385, 416
\bibitem[a]{} Roberts, D. H., Leh\'ar, J., Hewitt, J. N., and 
	Burke, B. F. 1991, Nature, 352, 43
\bibitem[a]{} Schild, R. 1990, \apj, 100, 1771
\bibitem[a]{} Schild, R. E., and Cholfin, B. 1986, \apj, 300, 209
\bibitem[a]{} Schild, R., and Thomson, D. J. 1995, \aj, 109, 1970
\bibitem[a]{} Serra-Ricart, M., Oscoz A., Sanch\'\i s, T., 
	Mediavilla, E., Goicoechea, L. J., Licandro, J., Alcalde, 
	D., and Gil-Merino, R. 1999, \apj, 526, 40
\bibitem[a]{} Vanderriest, C., Schneider, J., Herpe, G., 
	Chevreton, M., Moles, M., and Wl\'erick, F. 1989, A\&A, 
	215, 1
\bibitem[a]{} Walsh, D., Carswell, R. F., and Weymann, R. J. 1979, 
	Nature, 279, 381
\end{thebibliography}
\end{document}